# Thermoelectric Effect in Mott Variable-Range Hopping

Takahiro Yamamoto[1,2], Masao Ogata[3,4], and Hidetoshi Fukuyama[2]

[1] *Department of Physics, Tokyo University of Science, Kagurazaka 1-3, Shinjuku, Tokyo 162-8601, Japan*
[2] *RIST, Tokyo University of Science, Shinjuku, Tokyo 162-8601, Japan*
[3] *Department of Physics, The University of Tokyo, Hongo 7-3-1 Bunkyo, Tokyo 113-0033, Japan*
[4] *Trans-scale Quantum Science Institute, University of Tokyo, Hongo, Bunkyo-ku, Tokyo 113-0033, Japan*

On the basis of the Kubo–Luttinger linear response theory combined with the scaling theory of Anderson localization predicting the energy dependence of localization length near the mobility edge, we have studied the thermoelectric response of a disordered system exhibiting Mott variable-range hopping (VRH). We found that the low-tempetature Seebeck coefficient for VRH conduction in a $d$-dimensional system varies as $S \propto T^{d/(d+1)}$, which is different from the widely used expression, $S \propto T^{(d-1)/(d+1)}$, based on the energy-independent localization length. It is seen that the low-$T$ behavior of $S(T)$ of the thiospinel CuCrTiS$_4$ is in full agreement with the new scheme.

## 1. Introduction

Most thermoelectric (TE) materials with a high power factor and high efficiency are disordered semiconductors. The TE response in these strongly disordered systems cannot be described by the Boltzmann transport equation (BTE), and this was a long-standing problem in condensed matter physics. In linear response theory, the electrical current density $\boldsymbol{j}$ under the electric field $\boldsymbol{E}$ and the temperature gradient $\nabla T$ can be described as

$$\boldsymbol{j} = L_{11}\boldsymbol{E} - L_{12}\frac{\nabla T}{T}, \tag{1}$$

where $L_{11}$ is the electrical conductivity and $L_{12}$ is the thermoelectric conductivity. Recently, the authors have clarified that, on the basis of the Kubo–Luttinger linear response theory,[1,2] $L_{11}$ and $L_{12}$ can be expressed as the Sommerfeld-Bethe (SB) relation,

$$L_{11}(T) = \int_{-\infty}^{\infty} d\varepsilon \left(-\frac{\partial f_0}{\partial \varepsilon}\right) \sigma(\varepsilon, T), \tag{2}$$

$$L_{12}(T) = -\frac{1}{e}\int_{-\infty}^{\infty} d\varepsilon \left(-\frac{\partial f_0}{\partial \varepsilon}\right)(\varepsilon - \mu)\sigma(\varepsilon, T), \tag{3}$$

even for strongly disordered systems as long as the electron–phonon and electron mutual interactions are not importnt.[3,4] Here, $e$ is the elementary charge, $f_0$ is the Fermi–Dirac distribution function, $\mu$ is the chemical potential, and $\sigma(\varepsilon, T)$ is the spectral conductivity. Using $L_{11}$ and $L_{12}$, the Seebeck coefficient $S$ can be described as

$$S(T) = \frac{1}{T}\frac{L_{12}(T)}{L_{11}(T)}. \tag{4}$$

The SB relation makes it possible to analyze correctly the TE response of strongly disordered systems that cannot be treated by BTE (see, for example, Refs. 5–8).

Mott variable-range hopping (VRH) well describes low-$T$ conductivity, $L_{11}$, in strongly disordered systems with localized electronic states.[9,10] The conductivity of hopping distance $R$ is assumed as

$$L_{11} = \sigma_0^L \exp\left\{-2\alpha R - \frac{W}{k_B T}\right\}, \tag{5}$$

where $\alpha$ is the inverse of the localization length of the wave function and $W$ is the characteristic energy difference in hopping given by $W = 1/(R^d N_d)$, where $N_d$ is the number of localized states per unit volume in $d$ dimension. By optimizing $R$, we obtain the optimized distance

$$R_{\text{opt}} = \left(\frac{d}{2\alpha k_B T N_d}\right)^{\frac{1}{d+1}}, \tag{6}$$

the corresponding characteristic energy difference

$$W_{\text{opt}} = \frac{1}{N_d}\left(\frac{2\alpha k_B T N_d}{d}\right)^{\frac{d}{d+1}}, \tag{7}$$

and

$$L_{11} = \sigma_0^L \exp\left\{-\frac{d+1}{d}\left(\frac{d(2\alpha)^d}{k_B T N_d}\right)^{\frac{1}{d+1}}\right\}. \tag{8}$$

On the other hand, the $T$ dependence of $S(T)$ for VRH conduction is phenomenologically treated as follows.[10–14] The states near the Fermi energy within a characteristic energy width $\Delta$ will contribute to the TE effect for Mott VRH conduction, where $\Delta$ is estimated as $\Delta \sim W_{\text{opt}} \propto T^{d/(d+1)}$. As a result, the $T$ dependence of $S(T)$ in $d$ dimensions is estimated as

$$S \propto \Delta^2/T \propto T^{(d-1)/(d+1)}. \tag{9}$$

However, this argument does not take into account the energy dependence of the localization length near the mobility edge $\varepsilon_c$, where the localization length diverges, *i.e.*, $\xi(\varepsilon_c) = \infty$. In fact, several materials exhibiting Mott VRH conduction (*e.g.*, the thiospinel CuCrTiS$_4$) are known not to obey Eq. (9), and this issue remains to be elucidated. The main purpose of this work is to clarify the $T$ dependence of $S(T)$ for Mott VRH conduction.

## 2. Theoretical Scheme

In the following discussion, we assume n-type semiconductors unless otherwise noted. In the scaling theory of Anderson localization,[15,16] the $\varepsilon$ dependence of localization length $\xi(\varepsilon)$ near $\varepsilon = \varepsilon_c$ is given by

$$\xi(\varepsilon) = \xi_0 \left|\frac{\varepsilon_c - \varepsilon}{\varepsilon_c}\right|^{-s} \quad (\varepsilon < \varepsilon_c) \tag{10}$$

with a scaling dimension $s$. Therefore, the low-$T$ behavior of $S(T)$ obtained using Eq. (10) will be different from those obtained with Eq. (9).

We assume that the spectral conductivity for Mott VRH conduction in the localization region is given by

$$\sigma_L(\varepsilon, T) = \sigma_0^L \exp\left[-\frac{d+1}{d}\left(\frac{T_d(\varepsilon)}{T}\right)^{\frac{1}{d+1}}\right]. \tag{11}$$

Here, $T_d$ is the characteristic temperature given by

$$T_d = \frac{d}{k_B N_d \xi^d(\varepsilon)}, \tag{12}$$

where we have assumed $1/(2\alpha)$ in Eq. (8) to be $\xi(\varepsilon)$ given by Eq. (10). In Eq. (11), $\sigma_L^0$ is taken to be independent of $\varepsilon$ since the $\varepsilon$ dependence of $\xi(\varepsilon)$ is stronger. To emphasize the $\varepsilon$ dependence of $\sigma_L(\varepsilon, T)$, we rewrite Eq. (11) as

$$\sigma_L(\varepsilon, T) = \sigma_0^L \exp\left\{-A(T)\left|\frac{\varepsilon_c - \varepsilon}{\varepsilon_c}\right|^r\right\} \tag{13}$$

with $A(T) \equiv \frac{d+1}{d}\left(\frac{d}{k_B T N_d \xi_0^d}\right)^{\frac{1}{d+1}}$ and $r \equiv \frac{d}{d+1}s$.

In addition, the delocalized states for $\varepsilon > \varepsilon_c$ would affect essentially the $T$ dependence of $S(T)$ as $T$ increases, which is not taken into account in Mott–Davis's argument. According to the scaling theory, the spectral conductivity near $\varepsilon = \varepsilon_c$ in the delocalization region is known to be

$$\sigma_{DL}(\varepsilon, T) = \sigma_0^{DL} \left|\frac{\varepsilon - \varepsilon_c}{\varepsilon_c}\right|^{s'} \tag{14}$$

with a scaling dimension $s'$. Thus, for an n-type semiconductor, the total spectral conductivity is expressed as

$$\sigma(\varepsilon, T) = \sigma_L(\varepsilon, T)\theta(\varepsilon_c - \varepsilon) + \sigma_{DL}(\varepsilon, T)\theta(\varepsilon - \varepsilon_c), \tag{15}$$

where $\theta(x)$ is the Heaviside step function, i.e., $\theta(x > 0) = 1$ and $\theta(x < 0) = 0$.

By substituting Eq. (15) into Eqs. (2) and (3), we obtain

$$L_{11}(T) = L_{11}^L(T) + L_{11}^{DL}(T) \tag{16}$$

$$L_{12}(T) = -\frac{k_B T}{e}\{\mathcal{L}_{12}^L(T) + \mathcal{L}_{12}^{DL}(T)\} \tag{17}$$

with the contribution from localized states,

$$L_{11}^L(T) = \frac{\sigma_0^L}{4}\int_0^\infty dx \frac{1}{\cosh^2\frac{X-x}{2}}\exp\left\{-A(T)\left(\frac{k_B T}{\varepsilon_c}x\right)^r\right\}, \tag{18}$$

$$\mathcal{L}_{12}^L(T) = \frac{\sigma_0^L}{4}\int_0^\infty dx \frac{X-x}{\cosh^2\frac{X-x}{2}}\exp\left\{-A(T)\left(\frac{k_B T}{\varepsilon_c}x\right)^r\right\}, \tag{19}$$

and the contribution from delocalized states,

$$L_{11}^{DL}(T) = \frac{\sigma_0^{DL}}{4}\left(\frac{k_B T}{\varepsilon_c}\right)^{s'}\int_0^\infty dx \frac{x^{s'}}{\cosh^2\frac{X+x}{2}}, \tag{20}$$

$$\mathcal{L}_{12}^{DL}(T) = \frac{\sigma_0^{DL}}{4}\left(\frac{k_B T}{\varepsilon_c}\right)^{s'}\int_0^\infty dx \frac{(X+x)x^{s'}}{\cosh^2\frac{X+x}{2}}, \tag{21}$$

where $X \equiv \frac{\varepsilon_c - \mu}{k_B T} > 0$ corresponds to the localization region and $X < 0$ the delocalization one. Using Eqs. (16) and (17), the Seebeck coefficient in Eq. (4) is given by

$$S(T) = -\frac{k_B}{e}\frac{\mathcal{L}_{12}^L(T) + \mathcal{L}_{12}^{DL}(T)}{L_{11}^L(T) + L_{11}^{DL}(T)}. \tag{22}$$

Let us consider the limit of $T \to 0$ for the localization region, i.e., $X \gg 1$. In this limit, $L_{11}^L$ and $L_{11}^{DL}$ are respectively represented by

$$L_{11}^L(T) \approx \sigma_0^L \exp\left\{-A(T)\left(\frac{\varepsilon_c - \mu}{\epsilon_c}\right)^r\right\} \tag{23}$$

and

$$L_{11}^{DL}(T) \approx \sigma_0^{DL} e^{-X}\Gamma(1 + s')\left(\frac{k_B T}{\varepsilon_c}\right)^{s'}, \tag{24}$$

where $\Gamma(x)$ is the gamma function. Similarly, $\mathcal{L}_{12}^L$ and $\mathcal{L}_{12}^{DL}$ for $X \gg 1$ are approximately given by

$$\mathcal{L}_{12}^L(T) \approx \sigma_0^L \frac{r\pi^2}{3}\left(\frac{\varepsilon_c - \mu}{\varepsilon_c}\right)^{r-1}\frac{k_B T}{\varepsilon_c}A(T)$$

$$\times \exp\left\{-A(T)\left(\frac{\varepsilon_c - \mu}{\epsilon_c}\right)^r\right\} \tag{25}$$

and

$$\mathcal{L}_{12}^{DL}(T) \approx \sigma_0^{DL} X e^{-X}\Gamma(1 + s')\left(\frac{k_B T}{\varepsilon_c}\right)^{s'}. \tag{26}$$

In the low-$T$ limit $X \to \infty$, the contributions from the delocalized states, $L_{11}^{DL}(T)$ and $\mathcal{L}_{12}^{DL}(T)$, are negligible, and eventually, the low-$T$ Seebeck coefficient in Eq. (22) is given by

$$S(T) \approx -\frac{r\pi^2 k_B}{3e}\left(\frac{\varepsilon_c - \mu}{\varepsilon_c}\right)^{r-1}A(T)\frac{k_B T}{\varepsilon_c} \propto T^{d/(d+1)}, \tag{27}$$

which is different from the widely used $S \propto T^{(d-1)/(d+1)}$ in Eq. (9).

## 3. Comparison with Experiment

The thiospinel CuCrTiS$_4$ with high purity was synthesized by Kariya et al.[17] and has attracted considerble attention because it exhibits a large $S$ value at room temperature [$S(300\ \text{K}) = -180\ \mu\text{V/K}$].[18] Recently, Berthebaud et al. have reported that the thiospinel CuCrTiS$_4$ exhibits Mott VRH conduction when $d = 3$ at low $T$ below 75 K, but its Seebeck coefficient does not obey the Mott–Davis expression $S(T) \propto T^{1/2}$ in Eq. (9),[19] and the experimental data of $S(T)$ were fitted by $S = aT^{1/2} + bT$, where the $T$-linear term is an ad hoc term added to reproduce the experimental data. Thus, we attempt to reproduce the experimental data of the thiospinel CuCrTiS$_4$ by the present theoretical scheme without any ad hoc term.

Figures 1(a) and 1(b) represent the $T$-dependence of resistivity $\rho(T) = 1/L_{11}(T)$ and Seebeck coefficient $S(T)$ of the thiospinel CuCrTiS$_4$ at low $T$ below $T < 75$ K. In Figs. 1(a) and 1(b), the open circles are the experimental data in Ref. 19 and the solid curves are the theoretical ones calculated from Eqs. (16) and (19) using $\sigma_0^L$, $\sigma_0^{DL}$, $N_d\xi_0^d$, and $\mu(T)$ as fitting parameters. Here, the mobility edge $\varepsilon_c$ is taken to be $\varepsilon_c = 0.5$ eV as a typical value of characteristic energy, and the scaling dimensions $s$ and $s'$ are assumed to be $s = s' = 1/2$ based on Ref. 20. We now explain how to determine the fitting parameters $\sigma_0^L$, $\sigma_0^{DL}$, $N_d \xi_0^d$, and $\mu(T)$. First, $\sigma_0^L$, $N_d \xi_0^d$, and $\mu(0)$ are determined as $\sigma_0^L = 2.5 \times 10^5$ S/cm, $N_d \xi_0^d = 1.45 \times 10^{-2}$ (eV)$^{-1}$, and $\varepsilon_c - \mu(0) = 119$ meV by fitting the experimental data of $\rho(T)$ and $S(T)$ in the low-$T$ region obeying $X \gg 1$ using Eqs. (18) and (19). Next, by fitting all the experimental data of $\rho(T)$ shown in Fig. 1(a) using Eqs. (16) and (17), $\sigma_0^{DL}$ is determined as $\sigma_0^{DL} = 10^7$ S/cm and $\mu(T)$ is obtained as represented in Fig. 2.

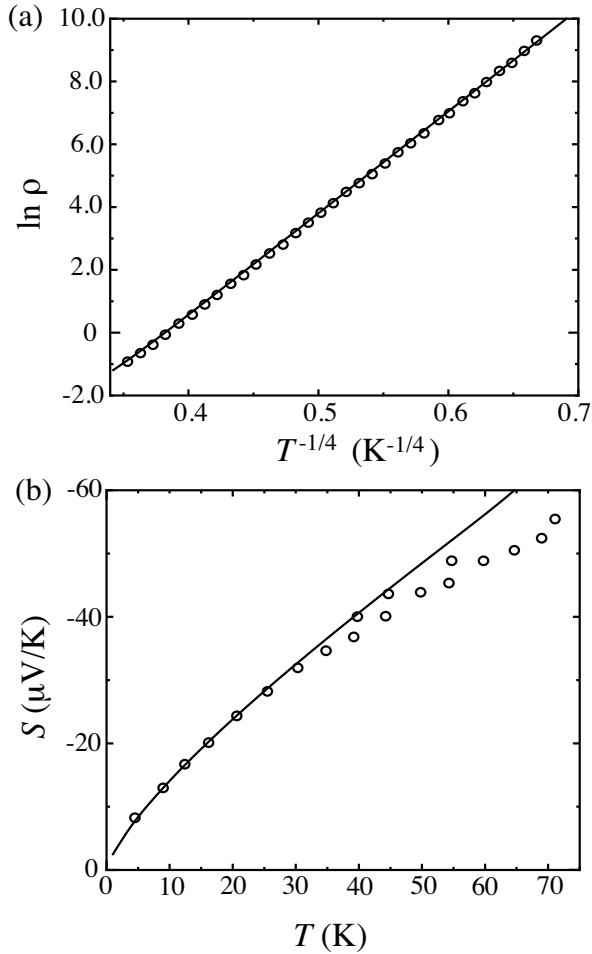

**Fig. 1.** Temperature dependence of the resistivity (a) and Seebeck coefficient (b) of the thiospinel CuCrTiS$_4$ below 75 K. The open circles are the experimental data in Ref. 19 and the solid curve is the theoretical value.

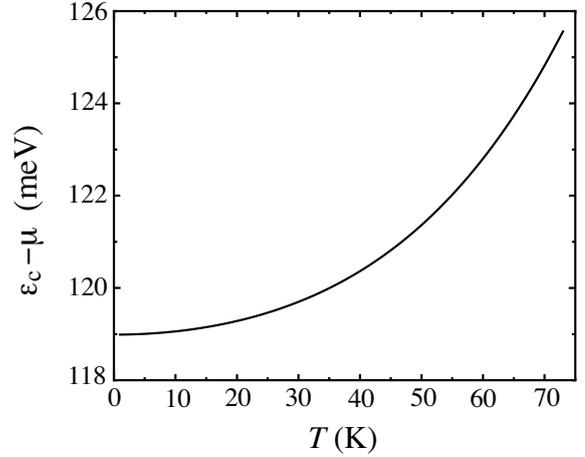

**Fig. 2.** Temperature dependence of the chemical potential $\mu(T)$ of the thiospinel CuCrTiS$_4$, which is obtained from fitting the experimental data in Figs. 1(a) and 1(b). Here, $\varepsilon_c$ is the mobility edge.

As a result of the above-mentioned fitting, the theoretical curve (solid curve) of $\rho(T)$ in Fig. 1(a) is in excellent agreement with the experimental data (open circles). Thus, we reproduced the experimental result that $\rho(T)$ for $X \gg 1$ exhibits as $\ln \rho(T) \propto T^{-1/4}$, which is the well-known behavior for Mott VRH conduction when $d = 3$. In addition, the theoretical curve (solid curve) of $S(T)$ in Fig. 1(b) is in good agreement with the experimental data represented by the open circles in the low-$T$ region below $T \lesssim 30$ K and it varies as $S \propto T^{3/4}$ as we proposed in Eq. (27), which is different from the Mott–Davis expression $S \propto T^{1/2}$ in Eq. (9) with $d = 3$. Thus, we succeeded in reproducing the experimental result by our theoretical scheme taking into account an energy-dependent localization length in Eq. (10). Noted that the Mott–Davis expression is expected to be valid when the chemical potential $\mu$ is far sufficiently from the mobility edge $\varepsilon_c$ and the energy dependence of $\xi(\varepsilon)$ is negligible.

Finally, we again emphasize that the Seebeck coefficient of the thiospinel CuCrTiS$_4$ in the low-$T$ region of $T \lesssim 30$ K is in full agreement with the present theoretical scheme. On the other hand, the theoretical curve of $S(T)$ in Fig. 1(b) deviates upward from the experimental data in the high-$T$ region of $T \gtrsim 30$ K. This implies the existence of some factors that are not included in the present scheme, which is left for future work.

## 4. Summary

In this work, we introduced a new theoretical scheme of the Seebeck coefficient for Mott VRH conduction based on the SB relation, taking into account the energy dependence of the localization length and the $T$-dependent chemical potential. On the basis of the new scheme, we found that the low-$T$ Seebeck coefficient for VRH conduction in a $d$-dimensional system varies as $S \propto T^{d/(d+1)}$. This is different from the widely used Mott–Davis expression, $S \propto T^{(d-1)/(d+1)}$. In addition, using the new scheme, we succeeded in reproducing the Seebeck coefficient of the thiospinel CuCrTiS$_4$ in the low-$T$ region below about 30 K. On the other hand, we could not reproduce the experimental data in the high-$T$ region of $T \gtrsim 30$ K. This implies the existence of some factors that are not incorporated in the present theoretical scheme, and this issue remains to be elucidated.

We also note the validity of the SB relation in Eq. (3) for VRH that involves electron hopping between localized states with the help of phonons. Although some processes associated with electron–phonon interactions cannot be described by the SB relation,[4] the essence of VRH is the energy transfer between electrons and phonons, which can be properly treated by the SB relation independently of the details of electron–phonon interactions. In this context, however, cases with coulomb interactions will need more careful study.

We thank K. Sasaoka, M. Matsubara, and K. Takahashi for fruitful discussions. This work was partly supported by JSPS KAKENHI (Grant Nos. 18H01816 and 18H01162), the JST CREST program (Grant No. JPMJCR17I5), and the JST MIRAI program (Grant No. JPMJMI19A1).